%% file: sample-sigconf-authordraft.tex
\documentclass[sigconf]{acmart}
\usepackage{xspace}
\usepackage{enumitem}
\usepackage{multirow}
\usepackage[normalem]{ulem}

\newcommand{\ie}{\emph{i.e., }}
\newcommand{\eg}{\emph{e.g., }}

\useunder{\uline}{\ul}{}
\newcommand{\ours}{TopKGAT\xspace}
\AtBeginDocument{%
  }

\copyrightyear{2026}
\acmYear{2026}
\setcopyright{cc}
\setcctype{by}
\acmConference[WWW '26] {Proceedings of the ACM Web Conference 2026}{April 13--17, 2026}{Dubai, United Arab Emirates.}
\acmBooktitle{Proceedings of the ACM Web Conference 2026 (WWW '26), April 13--17, 2026, Dubai, United Arab Emirates}
\acmISBN{979-8-4007-2307-0/2026/04}
\acmDOI{10.1145/3774904.3792717}
\begin{document}

\title{\ours: A Top-K Objective-Driven Architecture for Recommendation}


\author{Sirui Chen}
\orcid{0009-0006-5652-7970}
\affiliation{
    \institution{Zhejiang University} 
    \city{Hangzhou} 
    \country{China}}
\email{chenthree@zju.edu.cn}
\authornotemark[2]
\authornotemark[3]

\author{Jiawei Chen}
\orcid{0000-0002-4752-2629}
\affiliation{
    \institution{Zhejiang University} 
    \city{Hangzhou} 
    \country{China}}
\email{sleepyhunt@zju.edu.cn}
\authornote{Corresponding author.}
\authornote{State Key Laboratory of Blockchain and Data Security, Zhejiang University.}
\authornote{College of Computer Science and Technology, Zhejiang University.}
\authornote{Hangzhou High-Tech Zone (Binjiang) Institute of Blockchain and Data Security.}

\author{Canghong Jin}
\orcid{0000-0002-9774-9688}
\affiliation{
    \institution{Hangzhou City University} 
    \city{Hangzhou} 
    \country{China}}
\email{jinch@zucc.edu.cn}

\author{Sheng Zhou}
\orcid{0000-0003-3645-1041}
\affiliation{
    \institution{Zhejiang University} 
    \city{Hangzhou} 
    \country{China}}
\email{zhousheng_zju@zju.edu.cn}

\author{Jingbang Chen}
\orcid{0000-0002-7279-0801}
\affiliation{
    \institution{CUHK-Shenzhen \& SLAI}
    \city{Shenzhen} 
    \country{China}}
\email{chenjb@cuhk.edu.cn}

\author{Wujie Sun}
\orcid{0000-0001-7739-3517}
\affiliation{
    \institution{Zhejiang University} 
    \city{Hangzhou} 
    \country{China}}
\email{sunwujie@zju.edu.cn}

\author{Can Wang}
\orcid{0000-0002-5890-4307}
\affiliation{
    \institution{Zhejiang University} 
    \city{Hangzhou} 
    \country{China}}
\email{wcan@zju.edu.cn}
\authornotemark[2]
\authornotemark[4]

\renewcommand{\shortauthors}{Sirui Chen, et al.}

\begin{abstract}
\input{content/sec0-abstract}
\end{abstract}




\maketitle

\input{content/sec1-Introduction}
\begin{figure*}[!t]
    \centering
    \includegraphics[width=0.9\linewidth]{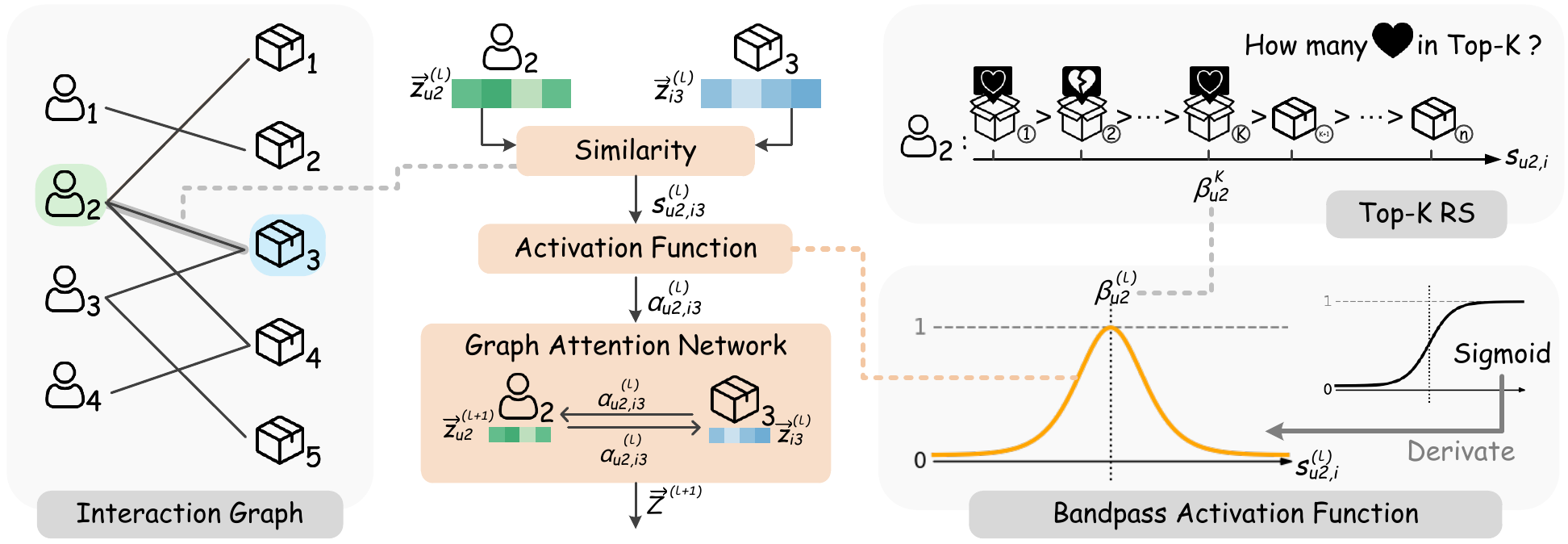}
    \caption{The illustration of \ours.}
    \label{fig: method}
    \Description{}
\end{figure*} 
\input{content/sec2-Preliminary}
\input{content/sec3-Methodology}

\input{content/sec4-Experiments}
\input{content/sec5-RelatedWork}
\input{content/sec6-Conclusions}

\begin{acks}
    \input{content/sec7-Acknowledgments}
\end{acks}

\bibliographystyle{ACM-Reference-Format}
\balance
\bibliography{content/main}
\appendix
\input{content/sec9-Appendices}

\end{document}

%% file: content/sec0-abstract.tex
Recommendation systems (RS) aim to retrieve the top‑K items most relevant to users, with metrics such as Precision@K and Recall@K commonly used to assess effectiveness. The architecture of an RS model acts as an inductive bias, shaping the patterns the model is inclined to learn. In recent years, numerous recommendation architectures have emerged, spanning traditional matrix factorization, deep neural networks, and graph neural networks. However, their designs are often not explicitly aligned with the top‑K objective, thereby limiting their effectiveness.

To address this limitation, we propose TopKGAT, a novel recommendation architecture directly derived from a differentiable approximation of top‑K metrics. The forward computation of a single TopKGAT layer is intrinsically aligned with the gradient ascent dynamics of the Precision@K metric, enabling the model to naturally improve top‑K recommendation accuracy. Structurally, TopKGAT resembles a graph attention network and can be implemented efficiently. Extensive experiments on four benchmark datasets demonstrate that TopKGAT consistently outperforms state‑of‑the‑art baselines.
The code is available at \url{https://github.com/StupidThree/TopKGAT}.

%% file: content/sec1-Introduction.tex
\section{Introduction}

Recommendation systems have become an essential infrastructure in modern online platforms, helping users discover relevant items from vast catalogs of products, videos, music, and other content \cite{isinkayeRecommendationSystemsPrinciples2015, koSurveyRecommendationSystems2022}. In practice, both display space and user attention are severely constrained --- whether on a mobile interface showing a handful of products, a streaming platform suggesting a few videos, or an e-commerce homepage featuring selected items. This constraint naturally formulates the \textit{top-K recommendation} problem. The primary objective is to maximize top-K metrics such as  $\mathrm{Precision}@K$, $\mathrm{Recall}@K$,  ensuring that the K displayed items are favored by the user, irrespective of those items outside the top-K positions. 

The architecture of an RS model plays a decisive role in determining its effectiveness. It is not merely an implementation detail but defines the patterns the model can learn and the \textit{inductive bias} it introduces during training \cite{chenBiasDebiasRecommender2023}. Over the past decade, various architectures for recommendation systems have been proposed, evolving from traditional matrix factorization methods \cite{korenMatrixFactorizationTechniques2009,korenAdvancesCollaborativeFiltering2011} to recent deep neural networks \cite{covingtonDeepNeuralNetworks2016,  zhangDeepLearningBased2019} and graph neural networks \cite{heLightGCNSimplifyingPowering2020, wuGraphNeuralNetworks2022}. Despite their success, many of these architectures are \emph{misaligned} with the top-K ranking goal, thereby limiting their effectiveness. For example, stacking multiple GNN layers can lead to over-smoothing, collapsing node representations, and reducing the model's ability to distinguish positive from negative instances --- conflicting with the top-K recommendation objective.

A notable exception is \emph{Rankformer} \cite{chenRankformerGraphTransformer2025}, which explicitly incorporates the ranking property of RS into its architectural design, achieving state-of-the-art performance. Its improvement stems from a specific neural network structure aligned with the gradient descent step of the Bayesian Personalized Ranking (BPR) objective, which promotes higher scores for positive instances relative to negatives. However, BPR optimizes pairwise comparisons across the entire item set, while the top-K objective (\eg Precision@K) focuses solely on the quality of the highest-ranked K items \cite{cremonesiPerformanceRecommenderAlgorithms2010}. This mismatch leaves a gap between BPR optimization and top-K performance. Although it adopts a linearization technique to approximate non-linear activations and a caching mechanism to store intermediate variables for acceleration, these strategies inevitably introduce approximation errors and incur significant memory consumption, thereby hindering its practical application. This raises an important research question: How can we design a recommendation architecture that is \emph{directly aligned} with the top-K objective? 

Towards this end, we propose \textbf{\ours}, a novel recommendation architecture derived directly from a differentiable approximation of top-K metrics. The key idea is to align the forward computation of a single-layer with the gradient ascent dynamics of the top-K objective, so that the architecture inherently drives improvements in top-K performance. Implementing this alignment is challenging because the top-K objective depends on ordinal item ranking positions and is inherently non-continuous. We address this by employing a quantile-based method \cite{haoQuantileRegression2007} to avoid explicit ranking computation and introducing a sigmoid-based smooth approximation \cite{yangBreakingTopKBarrier2025} to the Heaviside step function. This formulation yields a graph attention structure that aggregates information along user–item interaction edges, with attention weights determined by embedding similarity, while incorporating a custom activation function and a user-dependent bias term. The resulting architecture is concise and efficient,  fully aligned with the gradient of the top-K objective, and naturally enhances top-K recommendation accuracy. Extensive experiments on four benchmark datasets demonstrate that \ours consistently outperforms state-of-the-art baselines.

In summary, our main contributions are as follows:
\begin{itemize}[leftmargin=*,nosep]
\item We highlight the importance of integrating the top-K objective directly into designing the recommendation model architecture. 
\item We propose \ours, a novel recommendation model whose architecture is directly derived from the gradient of a top-K metric, thereby enhancing its capability in top-K recommendation.
\item We validate the effectiveness of \ours through experiments on four real-world datasets, showing consistent and substantial improvements over state-of-the-art baselines.
\end{itemize}

%% file: content/sec2-Preliminary.tex
\section{Preliminary}

In this section, we present the background of graph-based recommendation and graph attention networks.

\subsection{Top-K Recommendation} \label{sec: preliminary top-k recommendation}

\textbf{Problem Formulation.}
Following previous work \cite{wangNeuralGraphCollaborative2020,heLightGCNSimplifyingPowering2020, koSurveyRecommendationSystems2022}, we focus on the collaborative filtering scenario, the most general and widely studied setting in recommendation systems.
In a recommender system, there is a set of users $\mathcal{U}$ and a set of items $\mathcal{I}$. The number of users and items is $|\mathcal{U}| = n$ and $|\mathcal{I}| = m$, respectively. The historical interaction data is denoted as $\mathcal{D} = \{(u, i) | u \in \mathcal{U}, i \in \mathcal{I}\}$, where each element $(u, i) \in \mathcal{D}$ indicates that user $u$ interacted with item $i$ (\eg clicked, purchased, or rated positively).
For subsequent algorithm development, we further define the set of items that user $u$ has interacted with as $\mathcal{N}_u = \{i \in \mathcal{I} | (u, i) \in \mathcal{D}\}$  with cardinality $d_u = |\mathcal{N}_u|$, and similarly, the set of users who have interacted with item $i$ as $\mathcal{N}_i = \{u \in \mathcal{U} | (u, i) \in \mathcal{D}\}$ with cardinality $d_i = |\mathcal{N}_i|$.
In practical recommendation scenarios, systems typically present a limited number of items to users. Therefore, the top-K recommendation task aims to recommend $K$ items that each user is most likely to interact with based on the historical interaction data $\mathcal{D}$.

Modern recommendation models widely adopt the embedding-based paradigm, which has proven effective for capturing user preferences and item characteristics \cite{zhaoImprovingLatentFactor2015, bergGraphConvolutionalMatrix2017, koSurveyRecommendationSystems2022, linHowRecommendationModels2025}.
These models typically map each user $u$ and item $i$ into $d$-dimensional embeddings $\mathbf{z}_u, \mathbf{z}_i \in \mathbb{R}^d$ through various modeling approaches, so that all embedding vectors constitute a tensor $\mathbf {Z} \in \mathbb{R}^{(n+m)\times d}$.
The predicted preference score between user $u$ and item $i$ is then computed as the inner product $s_{ui} = \mathbf{z}_u^T \mathbf{z}_i$. The inner product naturally measures similarity in the embedding space and has been shown to be effective for collaborative filtering, widely used in existing work \cite{wangNeuralGraphCollaborative2020, heLightGCNSimplifyingPowering2020, songNGAT4RecNeighborawareGraph2021}.
For each user $u$, the system will recommend the top-K items with the highest predicted scores: $\mathcal{R}_u^K = \text{Top-K}_{i \in \mathcal{I} \setminus \mathcal{N}_u} \{s_{ui}\}$. 

\textbf{Graph-based Methods.} 
Graph-based recommendation models \cite{heLightGCNSimplifyingPowering2020,wuGraphNeuralNetworks2022, gaoSurveyGraphNeural2023} have emerged as a powerful approach for learning these embeddings by treating the user-item interactions as a bipartite graph $\mathcal{G} = (\mathcal{U} \cup \mathcal{I}, \mathcal{D})$. These models leverage graph neural networks to propagate information through the interaction structure, capturing collaborative signals that enhance recommendation quality. The embeddings are refined through multiple layers of graph neural networks, allowing the model to incorporate higher-order connectivity patterns in the user-item graph.

\textbf{Top-K Metrics.}
The fundamental principle of top-K metrics is to evaluate how well the recommended list $\mathcal{R}_u^K$ matches the user's true preferences reflected in $\mathcal{T}_u$, where $\mathcal{T}_u$ represents the ground-truth set of items that the user $u$ actually interacted with in the test set. 
Two of the most widely used top-K metrics are $\mathrm{Precision}@K$ and $\mathrm{Recall}@K$:
\begin{align}
\text{Precision}@K(u) = \frac{|\mathcal{R}_u^K \cap \mathcal{T}_u|}{K} \label{Pre@K} ;
\text{Recall}@K(u) = \frac{|\mathcal{R}_u^K \cap \mathcal{T}_u|}{|\mathcal{T}_u|}
\end{align}
where $\mathrm{Precision}@K$ measures the fraction of relevant items among the top-K recommendations, while $\mathrm{Recall}@K$ measures the fraction of relevant items that are successfully recommended.

These top-K metrics are particularly important in recommendation, because they reflect the practical constraints of the recommender system. By explicitly considering the top-K metrics in the model architecture, we can better model item and user representations that are more applicable to real-world recommendations.

\subsection{Graph Attention Networks}

Unlike traditional graph convolutional approaches that use fixed or structure-dependent weights, GAT \cite{velickovicGraphAttentionNetworks2017} learns to assign different importance to different neighbors, enabling more flexible and expressive representation learning, and has been widely adopted in various fields \cite{guoAttentionBasedSpatialtemporal2019, wangHeterogeneousGraphAttention2019,zhengGMANGraphMultiattention2020,wangKGATKnowledgeGraph2019,songSessionbasedSocialRecommendation2019}.

The core of GAT is its attention mechanism, which computes attention coefficients between connected nodes. For a target node $v$ and its neighbor $u \in \mathcal N_v$, the attention mechanism takes their features $\mathbf z_v,\mathbf z_u \in \mathbb R^d$ as inputs and computes: 
\begin{align}
& \alpha_{vu} = \text{softmax}_u(a(\mathbf{z}_v, \mathbf{z}_u)) \\
& \mathbf z_v'= \sum_{u \in \mathcal N_v}\alpha_{vu}\mathbf z_u
\end{align}
where $a$ is a function to calculate relevance score, and $\mathbf z_v'$ is the new representation of node $v$ after aggregation using GAT.


GAT is particularly well-suited for modeling entity representations in recommender systems. The attention mechanism naturally estimates the importance of different neighbors based on their features, which aligns with the fundamental principles of collaborative filtering that more similar users/items should have stronger influence.
NGAT4Rec \cite{songNGAT4RecNeighborawareGraph2021} represents the application of GAT to recommendation by adapting the attention mechanism for the collaborative filtering setting. It replaces the linear transformation-based relevance calculation in vanilla GAT with a dot-product similarity measure that is more natural for recommendation tasks, and substitutes Softmax normalization with degree-based normalization.

\subsection{Ranking-Inspired RS Models}

Traditional graph-based recommendation methods operate on the principle of making neighboring nodes more similar through iterative aggregation, essentially enhancing graph smoothness during the modeling process. However, this smoothness assumption does not align well with the ranking nature of recommendation tasks, where the goal is to discriminate between items rather than make them more similar. 
Recent RankFormer \cite{chenRankformerGraphTransformer2025} has attempted to address this misalignment by incorporating ranking objectives directly into neural architectures, proposing a graph transformer-based recommendation model.
By introducing specialized global attention mechanisms that capture pairwise ranking relationships, it can model representations that are inherently optimized for producing accurate rankings. This ranking-inspired design enables the model to better capture the comparative nature of user preferences.

Despite its innovations, RankFormer faces several limitations when applied to practical recommendation scenarios. 
First, while real-world systems focus on top-K recommendation where only a small subset of items is presented to users, RankFormer's inductive biases are designed for full ranking tasks. This mismatch means the model may spend computational resources optimizing for positions beyond the top-K that are irrelevant in practice.
Second, the transformer-based architecture introduces additional challenges. While the global receptive field allows the model to consider all user-item interactions simultaneously, this also dilutes the stronger signals from more relevant local neighbors, overshadowing the most reliable collaborative signals from directly connected neighbors with noise from distant nodes. Furthermore, the computational complexity of full attention necessitates oversimplified approximations, resulting in a deviation between what RankFormer actually optimizes and the ideal ranking objectives it was designed to capture. And even with these approximations, the time and space complexity of aggregating information from all nodes remains prohibitive for ultra-large-scale recommendation datasets. This scalability issue prevents its deployment in industrial recommendation systems where efficiency is paramount.

%% file: content/sec3-Methodology.tex
\section{Methodology}
In this section, we first introduce the architecture of our \ours, and then analyze its advantages in recommendation scenarios by comparing it with classic architectures such as GCN and GAT.

\subsection{\ours}

We aim to develop a graph neural network architecture that directly optimizes top-K recommendation metrics, bridging the gap between model design and evaluation objectives. 
The key challenge in optimizing top-K metrics lies in their discrete and non-differentiable nature, and top-K metrics are difficult to incorporate into the modeling process.
To address this challenge, we propose a two-step approach.
First, we transform the discrete top-K metric into a continuous and differentiable form by introducing quantile-based thresholds and smooth approximations of indicator functions.
Second, we design a graph attention network whose layer-wise updates directly follow the gradient direction of this differentiable metric, ensuring that the model's inductive bias aligns with top-K optimization objectives. 
For simplicity, here we focus on $\text{Precision}@K$ in our derivation, noting that $\text{Recall}@K$ differs only by a constant factor of $|\mathcal{T}_u|/K$.

\subsubsection{Top-K Metric Transformation.} 

The first challenge of optimizing the top-K metric is the discrete selection of top-K items, which involves a hard cutoff that is not amenable to gradient-based optimization. To address this, we refer to the quantile technique \cite{yangBreakingTopKBarrier2025} and introduce a quantile-based threshold $\beta_u^K$ that characterizes membership in the top-K set $\mathcal{R}_u^K$:
\begin{align}
\beta_u^K = \inf\{s_{ui} : i \in \mathcal{R}_u^K\}
\label{eq: beta}
\end{align}
This threshold allows us to rewrite the discrete set membership condition as a continuous comparison: an item $i$ belongs to the top-K recommendation set if and only if $s_{ui} \geq \beta_u^K$.

The second challenge arises from the indicator function in the metric computation, which is inherently discontinuous. Using the threshold-based characterization, we can express the intersection $|\mathcal{R}_u^K \cap \mathcal{T}_u|$ as:
\begin{align}
|\mathcal{R}_u^K \cap \mathcal{T}_u| = \sum_{i \in \mathcal{T}_u} \mathbb{I}(s_{ui} \geq \beta_u^K)
\end{align}
where $\mathbb{I}(\cdot)$ is the Heaviside step function. To make this expression differentiable, we replace the discontinuous indicator with the commonly used sigmoid function $\sigma(x)=1/(1+e^{-x})$, leading to our differentiable formulation of $\text{Precision}@K$:
\begin{align}
\text{Precision}@K(u) = \frac{\sum_{i \in \mathcal{T}_u} \sigma(s_{ui} - \beta_u^K)}{K}
\label{Pre@K-differentiable}
\end{align}

Through these transformations, we convert the discrete top-K metric $\text{Precision}@K$ into a continuous and differentiable objective.
This differentiable formulation serves as the foundation for designing our ranking-aware graph attention network.

\subsubsection{Model Architecture Design.} \label{sec: model architecture design}

Having obtained a differentiable top-K metric, we now design a graph neural network architecture that directly aligns with this objective. Our key insight is to simulate the gradient ascent optimization process within the network layers themselves, rather than relying solely on external loss functions. This approach offers two main advantages: first, it ensures that the model's inductive bias inherently favors top-K optimization throughout the embedding modeling process; second, we make the threshold $\beta_u^K$ learnable and layer-specific, denoted as $\beta_u^{(l)}$, enableing the model to progressively and adaptively refine its understanding of what constitutes a top-K recommendation for each user.

Following this design principle, we formulate the maximum optimization objective based on our differentiable $\text{Precision}@K$. 
Ignoring the constant denominator $K$ and adding L2 regularization for stability, we obtain:
\begin{align}
\mathcal{J}_{Pre@K} = \sum_{u \in \mathcal{U}} \sum_{i \in \mathcal{N}_u} \frac{\sigma(s_{ui} - \beta_u)}{\sqrt{d_ud_i}} - \lambda \|\mathbf{Z}\|_2^2
\label{eq: Loss-pre}
\end{align}
where $\lambda$ is the regularization coefficient. $\sqrt{d_ud_i}$ is a normalization coefficient to balance the contributions of users and items with different degrees, following existing graph-based RS methods \cite{heLightGCNSimplifyingPowering2020, songNGAT4RecNeighborawareGraph2021}.

To align the model architecture with this maximum objective, we simulate gradient ascent by updating embeddings according to:
\begin{align}
\mathbf{Z}^{(l+1)} = \mathbf{Z}^{(l)} + \tau \frac{\partial \mathcal{J}_{Pre@K}}{\partial \mathbf{Z}^{(l)}}
\label{eq: gradient descent}
\end{align}
where $\tau$ is the learning rate and $l$ denotes the layer index. By computing the gradients explicitly, we derive the aggregation formulas for \ours:
\begin{align}
\mathbf z_u^{(l+1)}=(1-\tau\lambda)\mathbf z_u^{(l)}+\tau\sum_{i \in \mathcal N_u}\frac{{\omega\left(({\mathbf z}_u^{(l)})^T{\mathbf z}_i^{(l)}-\beta_u^{(l)}\right)}}{\sqrt{d_ud_i}}\mathbf z_i^{(l)} \label{eq: after omega u} \\
\mathbf z_i^{(l+1)}=(1-\tau\lambda)\mathbf z_i^{(l)}+\tau\sum_{u \in \mathcal N_i}\frac{\omega\left(({\mathbf z}_u^{(l)})^T{\mathbf z}_i^{(l)}-\beta_u^{(l)}\right)}{\sqrt{d_ud_i}}\mathbf z_u^{(l)} \label{eq: after omega i} 
\end{align}
where $\omega(\cdot) = 4\sigma'(\cdot)=\frac{4}{(1+e^{-x})(1+e^{x})}$ is a scaled derivative of the sigmoid function. The detailed derivations are provided in Appendix \ref{app: derivation-layer}.

For simplicity and to reduce hyperparameters, we set $\lambda=\tau=1$, yielding:
\begin{align}
\mathbf z_u^{(l+1)}=\sum_{i \in \mathcal N_u}\frac{\omega\left(({\mathbf z}_u^{(l)})^T{\mathbf z}_i^{(l)}-\beta_u^{(l)}\right)}{\sqrt{d_ud_i}}\mathbf z_i^{(l)} \\
\mathbf z_i^{(l+1)}=\sum_{u \in \mathcal N_i}\frac{\omega\left(({\mathbf z}_u^{(l)})^T{\mathbf z}_i^{(l)}-\beta_u^{(l)}\right)}{\sqrt{d_ud_i}}\mathbf z_u^{(l)}
\label{eq: aggregation}
\end{align}
In practice, we normalize embeddings when computing similarities to ensure training stability.
The model maintains $L\times n$ learnable parameters $\beta_u^{(l)}$, allowing each user to have personalized relevance thresholds that evolve across layers.

The resulting architecture forms a graph attention network with a unique top-k ranking-aware design, where $({\mathbf z}_u^{(l)})^T{\mathbf z}_i^{(l)}$ measures the similarity between user and item embeddings, $\beta_u^{(l)}$ acts as a personalized threshold, and the weight function $\omega(\cdot)$ serves as a band-pass filter in the ranking space. It assigns maximum weights to user-item pairs whose similarities are near the decision boundary $\beta_u^{(l)}$, effectively focusing the model's attention on items that are critical for determining the top-K set. This design ensures that each layer of our network directly optimizes for top-K recommendation performance, making the model architecture inherently aligned with the evaluation metric.

\subsection{Model Analysis}

In this subsection, we analyze the key components of \ours and their relationships with existing graph-based recommendation methods. We demonstrate how our approach improves both GCN-based and GAT-based methods while introducing novel mechanisms specifically designed for top-K optimization.

\subsubsection{Relations with Graph-based Methods.}

\ours can be viewed as an improvement of existing graph-based recommendation methods, encompassing both GCN-based approaches like LightGCN \cite{heLightGCNSimplifyingPowering2020} and GAT-based methods like NGAT4Rec \cite{songNGAT4RecNeighborawareGraph2021}. 


LightGCN represents a simplified graph convolution approach where embeddings are updated through uniform neighbor aggregation.
It can be seen as a special case of our formulation where $\omega(\cdot) = 1$ (constant function) and $\beta_u^{(l)} = 0$. While the uniform aggregation of LightGCN is effective for general collaborative filtering, it treats all neighbors equally, regardless of their relevance to the top-K recommendation task.

Similarly, GAT-based methods like NGAT4Rec employ attention mechanisms with monotonic activation functions. NGAT4Rec can be viewed as another special case where $\omega(\cdot) = \text{ReLU}(\cdot)$ and $\beta_u^{(l)} = 0$. The ReLU activation assigns higher weights to items with higher similarity scores, following the intuition that more similar items should contribute more to the aggregation. However, this monotonic weighting may over-emphasize already high-scoring items that are confidently in the top-K set, while neglecting items near the decision boundary that are critical for optimizing the top-K recommendation result.

In contrast, \ours introduces two key innovations that make it particularly suited for top-K recommendation. First, our band-pass activation function $\omega(\cdot)$ focuses attention on items near the ranking boundary where improvements have the greatest impact on top-K metrics. Second, our learnable thresholds $\beta_u^{(l)}$ allow the model to adaptively determine the relevance boundary for each user at each layer, providing personalized and dynamic attention patterns. These design choices ensure that our architecture directly aligns with top-K optimization objectives, rather than relying on indirect optimization through rating prediction \cite{wangNeuralGraphCollaborative2020, heLightGCNSimplifyingPowering2020} or pairwise ranking losses \cite{cremonesiPerformanceRecommenderAlgorithms2010, yangPSLRethinkingImproving2024, zhangAdvancingLossFunctions2025} as in existing methods.

\subsubsection{Bandpass Activation Function.}

A distinguishing feature of \ours is the activation function $\omega(\cdot) = 4\sigma'(\cdot) = \frac{4}{(1+e^{-x})(1+e^{x})}$, which acts as a band-pass filter in the ranking space. Unlike traditional activation functions that are monotonically increasing (\eg ReLU, sigmoid, softmax), our activation function exhibits a bell-shaped curve centered at zero, as illustrated in Figure \ref{fig: method}.

This band-pass characteristic has profound implications for top-K recommendation. When applied to the shifted similarity score, the function assigns maximum weights to items whose scores are near the threshold $\beta_u^{(l)}$. Items with very high scores (confidently in the top-K set) or very low scores (clearly outside the top-K set) receive smaller weights, as their ranking positions are already well-established and contribute less to improving $\text{Precision}@K$.

From a signal processing perspective, this activation function filters out both high-frequency noise (random fluctuations among clearly irrelevant items) and low-frequency components (stable rankings of highly relevant items), focusing the model's learning capacity on the mid-frequency band where top-K decisions are most uncertain and impactful. This selective attention mechanism ensures efficient use of model capacity by concentrating on the most informative parts of the ranking distribution.

Furthermore, the specific form of $\omega(\cdot)$ as the derivative of the sigmoid function is not arbitrary. It emerges naturally from our gradient-based derivation. This mathematical grounding ensures that each layer's update directly follows the gradient direction of the differentiable $\text{Precision}@K$, making the band-pass filtering an inherent consequence of top-K optimization rather than an ad-hoc design choice.

\subsubsection{Learnable Top-K Threshold.}

The threshold parameter $\beta_u^{(l)}$ plays a crucial role in determining which part of the items in the ranked recommendation list receives the most attention during aggregation. While one could use the actual K-quantile of the score distribution as defined in Eq. \eqref{eq: beta}, we instead treat $\beta_u^{(l)}$ as learnable parameters. This design choice offers several important advantages.

First, computing exact quantiles during training would require sorting operations that introduce two critical issues: computational complexity and non-differentiability. Sorting has $O(nm \log m)$ complexity for each epoch, which becomes prohibitive when performed repeatedly during training. More fundamentally, the sorting operation is non-differentiable, making it incompatible with gradient-based optimization. Learnable thresholds elegantly sidestep both issues, allowing efficient computation and smooth gradient flow through standard backpropagation.

Second, learnable thresholds provide a flexible mechanism to capture hierarchical and personalized preference patterns. By making $\beta_u^{(l)}$ both user-specific and layer-specific, the model can learn sophisticated ranking strategies that vary across users and evolve through network depth. This dual adaptability enables the model to learn more nuanced representations that reflect both individual user characteristics and the multi-scale nature of preference formation, as we demonstrate in Section \ref{sec: case-study}.

Finally, learnable thresholds enable the model to adapt its focus throughout the training process. As we demonstrate in Section \ref{sec: case-study}, analyzing the learned threshold values reveals interesting patterns about user preferences and the model's ranking strategy across different layers and training stages.

%% file: content/sec4-Experiments.tex
\begin{table}[!t]
\centering
\caption{Statistics of datasets}
\label{tab: dataset}
\begin{tabular}{@{}lccc@{}}
\toprule
Dataset & \#User & \multicolumn{1}{l}{\#Item} & \multicolumn{1}{l}{\#Interaction} \\ \midrule
Ali-Display & 17,730 & 10,036 & 173,111 \\
Epinions & 17,893 & 17,659 & 301,378 \\
Food & 14,382 & 31,288 & 456,925 \\
Gowalla & 55,833 & 118,744 & 1,753,362 \\ \bottomrule
\end{tabular}%
\end{table}

\begin{table*}[!t]
\centering
\caption{Performance comparison between \ours and baselines. The best result is bolded and the runner-up is underlined. The mark `*' suggests the improvement is statistically significant with p < 0.05.}
\label{tab: comparison}
\resizebox{\textwidth}{!}{
\begin{tabular}{@{}cccccccccc@{}}
\toprule
 &  & \multicolumn{2}{c}{Ali-Display} & \multicolumn{2}{c}{Epinions} & \multicolumn{2}{c}{Food} & \multicolumn{2}{c}{Gowalla} \\ 
 &  & ndcg@20 & recall@20 & ndcg@20 & recall@20 & ndcg@20 & recall@20 & ndcg@20 & recall@20 \\ \midrule
\multirow{4}{*}{\begin{tabular}[c]{@{}c@{}}Non-attention\\ Methods\end{tabular}} & MF & 0.0606 & 0.1114 & 0.0518 & 0.0853 & 0.0186 & 0.0317 & 0.0935 & 0.1334 \\
 & LightGCN{[}SIGIR'20{]} & 0.0640 & 0.1201 & 0.0530 & 0.0881 & 0.0301 & {\ul 0.0499} & 0.1129 & 0.1594 \\
 & LightGCN++{[}RecSys'24{]} & 0.0571 & 0.1058 & {\ul 0.0566} & 0.0915 & 0.0299 & 0.0479 & {\ul 0.1175} & {\ul 0.1642} \\
 & ReducedGCN{[}PAKDD'25{]} & {\ul 0.0654} & {\ul 0.1216} & 0.0550 & {\ul 0.0922} & {\ul 0.0303} & 0.0490 & 0.1136 & 0.1600 \\ \midrule
\multirow{4}{*}{\begin{tabular}[c]{@{}c@{}}Attention-based\\ Methods\end{tabular}} & GAT & 0.0472 & 0.0881 & 0.0427 & 0.0727 & 0.0236 & 0.0390 & 0.0846 & 0.1246 \\ 
 & NGAT4Rec{[}Arxiv'20{]} & 0.0631 & 0.1188 & 0.0540 & 0.0888 & 0.0290 & 0.0480 & 0.1078 & 0.1524 \\
 & MGFormer{[}SIGIR'24{]} & 0.0649 & 0.1083 & 0.0542 & 0.0854 & 0.0260 & 0.0394 & 0.0973 & 0.1306 \\
 & Rankformer{[}WWW'25{]} & 0.0652 & 0.1208 & 0.0554 & 0.0895 & 0.0247 & 0.0424 & 0.1093 & 0.1589 \\ \midrule
\multirow{2}{*}{Our Method} & \multirow{2}{*}{\ours} & \textbf{0.0689*} & \textbf{0.1266*} & \textbf{0.0592*} & \textbf{0.0962*} & \textbf{0.0312*} & \textbf{0.0508*} & \textbf{0.1189*} & \textbf{0.1660*} \\
\multicolumn{2}{c}{} & 5.33\% & 4.10\% & 4.51\% & 4.32\% & 3.09\% & 1.80\% & 1.19\% & 1.13\% \\ \bottomrule
\end{tabular}
}
\end{table*}

\section{Experiments}

In this section, we conduct comprehensive experiments to answer the following research questions:
\begin{itemize}[leftmargin=*,nosep]
    \item \textbf{RQ1:} How does \ours perform compared to existing state-of-the-art methods?
    \item \textbf{RQ2:} What are the impacts of the important components (\ie benchmark term $\beta$ and activation function $\omega(\cdot)$) on \ours? 
    \item \textbf{RQ3:} How do the hyperparameters affect the performance of \ours?
    \item \textbf{RQ4:} How do different layers of \ours jointly optimize the ranking task in recommendation?
\end{itemize}

\subsection{Experimental Settings}

\subsubsection{Datasets.}
We conduct experiments on four real-world datasets: \textbf{Ali-Display}\footnote{https://tianchi.aliyun.com/dataset/dataDetail?dataId=56}, a dataset provided by Alibaba estimating click-through rates of Taobao display ads; \textbf{Epinions} \cite{caiSPMCSociallyawarePersonalized2017, zhaoImprovingLatentFactor2015}, a dataset collected from the online consumer review website Epinions.com; \textbf{Food} \cite{majumderGeneratingPersonalizedRecipes2019}, a user rating dataset collected from the recipe website Food.com; and \textbf{Gowalla} \cite{choFriendshipMobilityUser2011}, a dataset counting user check-ins on a location-based social platform. We use a standard 5-core setup and randomly split the dataset into training, validation, and test sets in a ratio of 7:1:2. The statistics of the datasets are shown in Table \ref{tab: dataset}.

\subsubsection{Metrics.}
We adopt two widely used metrics, $\text{Recall}@K$ and $\text{NDCG}@K$, to evaluate the recommendation accuracy, and refer to most studies on graph-based recommendation systems \cite{heLightGCNSimplifyingPowering2020, leeRevisitingLightGCNUnexpected2024, chenRankformerGraphTransformer2025} to simply set $K$ to $20$.

\subsubsection{Baselines.}

\textbf{1) Non-attention Methods.} 
The following four representative recommendation methods without attention mechanisms are included:
\begin{itemize}[leftmargin=*,nosep]
\item \textbf{MF} \cite{korenMatrixFactorizationTechniques2009}: A classical collaborative filtering approach that learns user and item embeddings through factorizing the interaction matrix without utilizing graph structure.
\item \textbf{LightGCN} \cite{heLightGCNSimplifyingPowering2020}: A simplified graph convolution method that removes feature transformation and nonlinear activation for recommendation.
\item \textbf{LightGCN++} \cite{leeRevisitingLightGCNUnexpected2024}:  An enhanced version of LightGCN that incorporates layer combination and embedding normalization to improve recommendation performance.
\item \textbf{ReducedGCN} \cite{kimReducedGCNLearningAdapt2025}: A recent variant of LightGCN that weakens irrelevant interactions through macro-scale neighborhood reduction and micro-scale edge weight reduction.
\end{itemize}

\textbf{2) Attention-based Methods.} 
The following four representative recommendation methods are included:
\begin{itemize}[leftmargin=*,nosep]
\item \textbf{GAT} \cite{velickovicGraphAttentionNetworks2017}: Applies learnable attention mechanisms to adaptively aggregate neighbor embeddings based on their feature similarities in the graph structure.
\item \textbf{NGAT4Rec} \cite{songNGAT4RecNeighborawareGraph2021} A neighbor-aware graph attention network for recommendation, which incorporates neighbor sampling and similarity calculation based on the vanilla GAT.
\item \textbf{MGFormer} \cite{chenMaskedGraphTransformer2024}: A multi-granularity transformer-based approach that captures both local graph structure and global collaborative signals through hierarchical attention mechanisms.
\item \textbf{Rankformer} \cite{chenRankformerGraphTransformer2025}: A ranking-aware graph transformer that explicitly models the ranking relationships between items through rank-sensitive attention.
\end{itemize}

For the compared methods, we use the source code provided officially and searched for optimal hyperparameters according to the instructions in the original paper. We extensively traversed the hyperparameter space as recommended by the authors to ensure that all compared methods achieved optimal performance.

\begin{table*}[t]
\centering
\caption{The result of the ablation study. The following table shows the ablation results after replacing the modules in \ours with the corresponding modules in the vanilla GAT.}
\label{tab: Ablation}
\resizebox{\textwidth}{!}{
\begin{tabular}{@{}c|cc|cccccccc@{}}
\toprule
 & \multirow{2}{*}{\begin{tabular}[c]{@{}c@{}}Threshold\\ Term?\end{tabular}} & \multirow{2}{*}{\begin{tabular}[c]{@{}c@{}}Activation\\ Function?\end{tabular}} & \multicolumn{2}{c}{Ali-Display} & \multicolumn{2}{c}{Epinions} & \multicolumn{2}{c}{Food} & \multicolumn{2}{c}{Gowalla} \\
 &  &  & ndcg@20 & recall@20 & ndcg@20 & recall@20 & ndcg@20 & recall@20 & ndcg@20 & recall@20 \\ \midrule
\ours-w/o-$\omega$\&$\beta$ & &  & 0.0512 & 0.0998 & 0.0313 & 0.0561 & 0.0105 & 0.0186 & 0.0648 & 0.1047 \\
\ours-w/o-$\omega$ & $\checkmark$ &  & 0.0493 & 0.1010 & 0.0291 & 0.0532 & 0.0163 & 0.0273 & 0.0593 & 0.0942 \\
\ours-w/o-$\beta$ & & $\checkmark$ & 0.0664 & 0.1243 & 0.0531 & 0.0882 & 0.0304 & 0.0494 & 0.1181 & 0.1648 \\
\ours & $\checkmark$ &$\checkmark$ & \textbf{0.0689} & \textbf{0.1266} & \textbf{0.0592} & \textbf{0.0962} & \textbf{0.0312} &\textbf{0.0508} & \textbf{0.1189} & \textbf{0.1660} \\ \midrule
\end{tabular}
}
\end{table*}

\begin{figure*}[!t]
    \centering
    \includegraphics[width=0.9\linewidth]{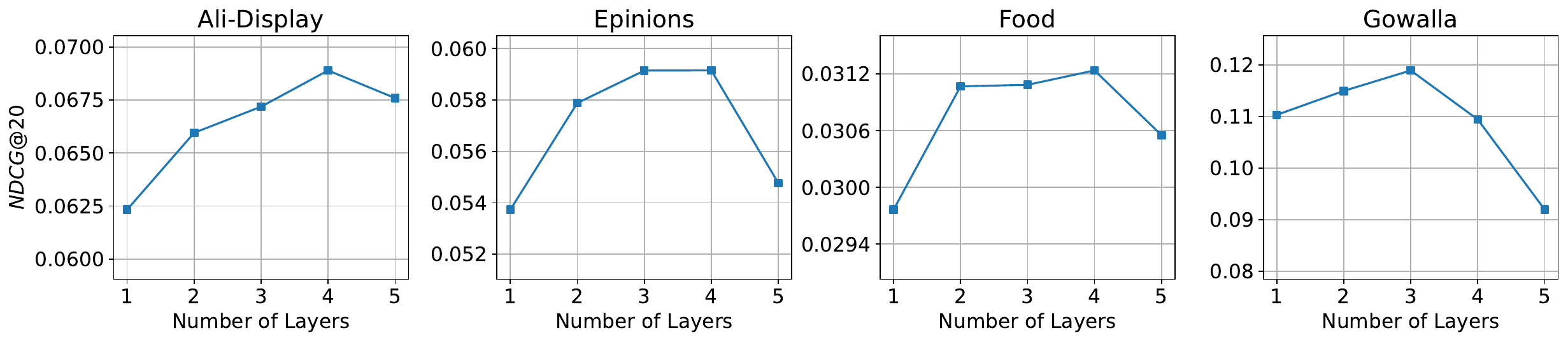}
    \caption{Performance of \ours in terms of $NDCG@20$ with different number of layers $L$.}
    \label{fig: layer}
    \Description{}
\end{figure*} 

\subsection{Performance Comparison (RQ1)}

The performance of our \ours compared to all baseline models in terms of $\text{Recall}@20$ and $\text{NDCG}@20$ is shown in Table \ref{tab: comparison}. Overall, \ours outperforms all comparison methods on all datasets, with an average improvement of 3.53\% in $\mathrm{NDCG}@K$ and 2.84\% in $\mathrm{Recall}@K$. This result demonstrates the effectiveness of our method.

\textbf{Compared with Non-attention Methods.}
\ours consistently outperforms all non-attention baselines across all datasets, demonstrating the effectiveness of incorporating attention mechanisms into graph-based recommendation. Unlike non-attention methods that employ fixed aggregation weights (\eg LightGCN's normalized sum pooling), \ours adaptively learns the importance of different neighbors through attention weights that are directly optimized for top-K ranking objectives. This adaptive aggregation allows our model to distinguish between neighbors with varying relevance to the target user's preferences, capturing more nuanced collaborative signals. 

\textbf{Compared with Attention-based Methods.}
\ours achieves superior performance compared to all attention-based baselines across all evaluation datasets. Notably, vanilla GAT performs worse than even the basic MF model on most datasets, indicating that the original GAT architecture is ill-suited for recommendation tasks without proper adaptation. In contrast, our method successfully adapts the GAT framework by explicitly connecting the attention mechanism to top-K optimization objectives, resulting in a recommendation-specific GAT architecture. MGFormer and RankFormer, which employ transformer architectures with global attention mechanisms, achieve performance comparable to state-of-the-art GCN-based methods on some datasets, while underperforming LightGCN on others. Their global attention mechanisms expand the receptive field during representation aggregation, but also introduce noise from less relevant nodes. 
Constrained by computational complexity, they often employ significant approximations, leading to reduced computational precision for important nodes compared to non-global aggregation methods.
Existing GAT-based methods like NGAT4Rec still employ attention mechanisms designed for general graph tasks, thus maintaining their focus on graph smoothness as the optimization objective, achieving performance similar to LightGCN.

\subsection{Ablation Study (RQ2)}
To investigate the contribution of each component in our framework, we conduct ablation studies by replacing the modules in \ours with the corresponding modules in the vanilla GAT, as shown in Table \ref{tab: Ablation}. Specifically, for the threshold term $\beta_u^{(l)}$, we set $\beta = 0$ in the removal experiments; for the activation function $\omega(\cdot)$, we replace our sigmoid-derivative-based function with GAT's softmax function. 

\textbf{Impact of Activation Function.}
The results demonstrate that removing our activation function $\omega(\cdot)$ a substantial decline in performance degradation across all datasets. This confirms that the softmax function employed in vanilla GAT is fundamentally incompatible with recommendation tasks. The sigmoid-derivative-based activation used by \ours directly corresponds to the gradient of $\mathrm{Precision}@K$, allowing the aggregated weights passed through the activation function to better optimize the top-K ranking objective for recommendation tasks.

\textbf{Interaction between Threshold and Activation.}
An interesting observation is that the threshold term $\beta_u^{(l)}$ only improves performance when combined with our activation function $\omega(\cdot)$. 
Comparing \ours-w/o-$\omega$\&$\beta$ with \ours-w/o-$\omega$, the threshold term becomes ineffective or even detrimental. This is expected because the threshold term is derived from our $\textrm{Precision}@K$ optimization framework and designed to work in conjunction with the sigmoid-derivative activation. When the activation function is misaligned with top-K objectives, the threshold term loses its theoretical foundation and cannot provide meaningful adaptive filtering of neighbors. This interdependence validates our unified framework, where both components work synergistically to optimize recommendation performance.

\subsection{Role of the parameters (RQ3)}


We investigate the effect of the number of layers $L$ in TopKGAT on recommendation performance. Figure \ref{fig: layer} shows the model performance across different numbers of layers on all four datasets. A consistent pattern emerges: performance initially improves with increasing the number of layers, then gradually degrades with further depth.

With shallow architectures, the model captures only immediate collaborative signals from direct user-item interactions and their low-hop neighbors, which limits the exploitation of higher-order connectivity patterns. As we increase the number of layers, the model benefits from expanded receptive fields that incorporate multi-hop collaborative signals, enabling a better understanding of complex user preferences through indirect connections. However, excessively deep architectures suffer from over-smoothing, where node representations become increasingly similar as information propagates through multiple layers. Additionally, deeper models aggregate information from exponentially larger neighborhoods, introducing noise from less relevant nodes that dilutes the signal from truly influential user-item interactions. Furthermore, the number of $\beta$ parameters also increases with larger $L$, making model training more difficult.

\subsection{Case Study (RQ4)} \label{sec: case-study}



\begin{figure*}[t]
    \centering
    \includegraphics[width=0.9\linewidth]{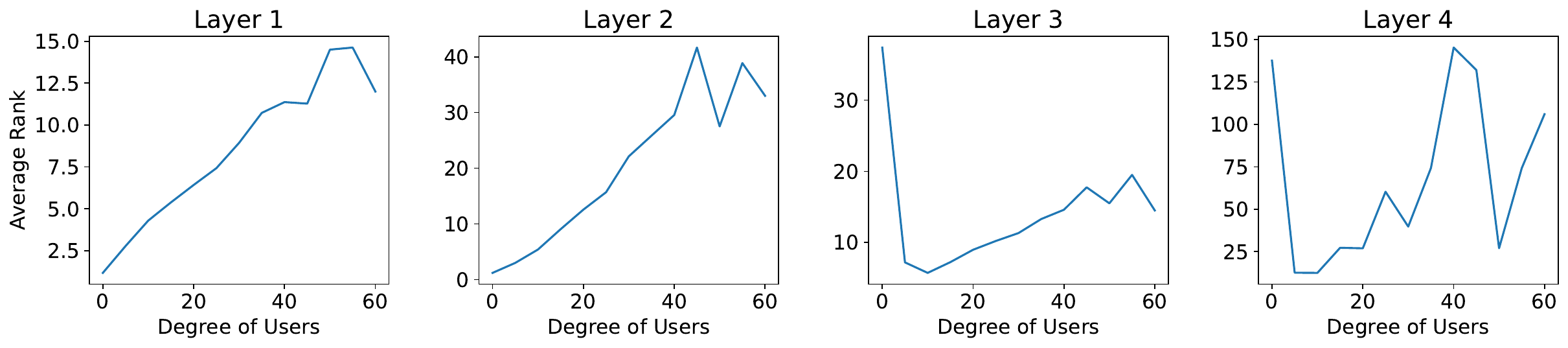}
    \caption{With different degrees of users, the average ranking of $\beta_u^{(l)}$ among $s_{ui}^{(l)}$ after training on Ali-Display.}
    \label{fig: beta-degree}
    \Description{}
\end{figure*} 

\begin{figure}[t]
    \centering
    \includegraphics[width=0.85\linewidth]{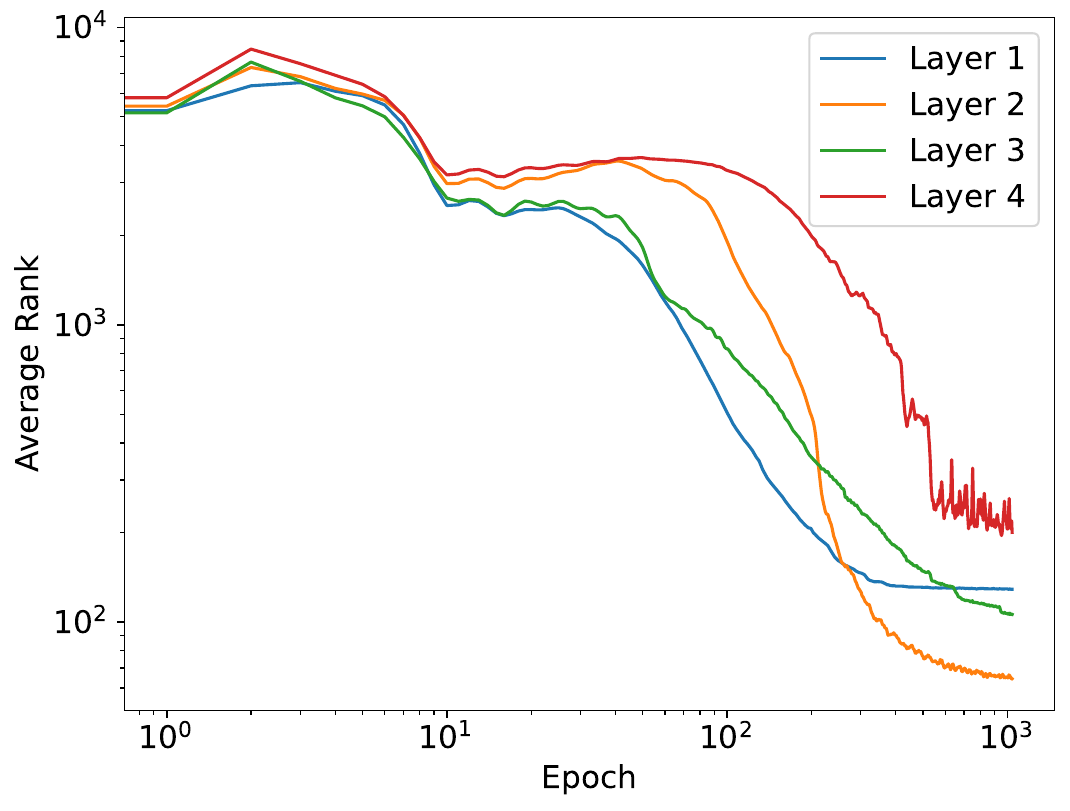}
    \caption{The average ranking of $\beta_u^{(l)}$ among $s_{ui}^{(l)}$ during training on the Ali-Display, plotted on a logarithmic axis.}
    \label{fig: beta-process}
    \Description{}
\end{figure} 

To gain deeper insights into how \ours learns to adaptively filter neighbors, we conduct a case study analyzing the learned threshold values $\beta_u^{(l)}$ across different layers and user characteristics, shown in Figures \ref{fig: beta-degree} and \ref{fig: beta-process}. For each user $u$ and layer $l$, we compute the ranking of $\beta_u^{(l)}$ among all item scores $s_{ui}^{(l)} = (\mathbf{z}_u^{(l)})^T\mathbf{z}_i^{(l)}$, which corresponds to the value of $K$ in the definition of $\beta_u^K$ in Eq.\eqref{eq: beta}. The ranking $K$ indicates that, within this layer, the model primarily optimizes items ranked near $K$ in the recommended list for user $u$.

Figure \ref{fig: beta-degree} shows how the average ranking of $\beta_u^{(l)}$ varies across layer $l$. As can be seen from the range of the vertical axis, shallow layers maintain smaller $K$ values, indicating they focus on optimizing top-ranked items, while deeper layers correspond to larger $K$ values, indicating that they focus on lower-ranked items. This hierarchical optimization strategy is intuitive: shallow layers establish strong representations by focusing on the most relevant items, while deeper layers further optimize the ranking by considering uncertain items, helping to better distinguish between relevant and irrelevant items in the middle range of the ranking.

Figure \ref{fig: beta-degree} also shows the relationship between the degree (\ie the number of items interacted with) of users in the training set and the ranking of the threshold.
In shallow layers, users with higher degrees tend to have larger $K$ values, suggesting that for active users, even shallow layers need to consider items beyond the very top rankings to effectively learn preferences from their rich interaction history. However, this correlation becomes more complex in deeper layers. In deeper layers, some users with higher degrees may also focus on higher-ranked positions with smaller $K$. This may help avoid being overwhelmed by the exponentially growing neighborhood information, maintain the quality of top-K recommendations. This adaptive behavior demonstrates \ours's ability to automatically adjust its optimization focus based on both user characteristics and layer depth.

Figure \ref{fig: beta-process} shows the evolution of the ranking of $\beta$ over time during training. The $K$ values of all layers consistently decrease. It suggests that as recommendation capabilities improve, the model shifts its focus of optimization toward higher-ranked positions in the recommendation list, directly focusing on optimizing the top-ranked positions. This aligns with the ultimate goal of top-K recommendations.

%% file: content/sec5-RelatedWork.tex
\section{Related Work}

\subsection{Graph-based Recommendation Systems}

The evolution of recommendation system architectures has witnessed several paradigm shifts: from early matrix factorization methods that model user-item interactions in latent spaces \cite{korenMatrixFactorizationTechniques2009,korenAdvancesCollaborativeFiltering2011}, to deep learning approaches that capture non-linear patterns \cite{covingtonDeepNeuralNetworks2016,  zhangDeepLearningBased2019}, and more recently to graph-based methods that explicitly model interaction structures \cite{heLightGCNSimplifyingPowering2020, wuGraphNeuralNetworks2022}, knowledge-graph enhanced systems that incorporate external information \cite{wangKGATKnowledgeGraph2019,wangKnowledgeGraphConvolutional2019}, and emerging LLM-based recommenders that leverage pre-trained language understanding 
\cite{cuiDistillationMattersEmpowering2024,wangMSLNotAll2025,cuiHatLLMHierarchicalAttention2025, cuiFieldMattersLightweight2025, wangLLM4DSRLeveragingLarge2025}
. Among all these approaches, graph-based methods have gained particular prominence due to their natural ability to model the inherently relational nature of user-item interactions and capture collaborative signals through graph structure without relying on external knowledge.

The development of graph neural networks for recommendation began with adapting general GNN architectures to bipartite user-item graphs. 
Early work, such as GCMC \cite{bergGraphConvolutionalMatrix2017} and NGCF \cite{wangNeuralGraphCollaborative2020}, applies graph convolutional networks to collaborative filtering. LightGCN \cite{heLightGCNSimplifyingPowering2020} demonstrates that the complex feature transformations and non-linearities in NGCF are unnecessary, achieving superior performance with a simplified architecture that only performs neighborhood aggregation. Subsequent work further makes various improvements based on LightGCN, such as introducing contrastive learning strategies \cite{caiLightGCLSimpleEffective2023, yuXSimGCLExtremelySimple2023}, improving graph structure \cite{wangCollaborationAwareGraphConvolutional2023, kimReducedGCNLearningAdapt2025}, and further simplifying GCN \cite{maoUltraGCNUltraSimplification2021, pengSVDGCNSimplifiedGraph2022}. 
However, these GCN-based methods fundamentally optimize for graph smoothness, encouraging similar representations for connected nodes. They inherently conflict with the discriminative requirements of recommendation, where the model must sharply distinguish between relevant and irrelevant items. Moreover, they suffer from well-known GNN limitations such as over-smoothing \cite{chenMeasuringRelievingOversmoothing2020}, over-squashing \cite{alonBottleneckGraphNeural2020}, and limitations in expressive power \cite{xuHowPowerfulAre2019}.

Introducing attention mechanisms into graph-based recommendation systems holds promise for addressing these limitations through adaptive, relevance-aware neighborhood information aggregation patterns.
A few attention-based graph recommendation methods adopt the GAT architecture, such as NGAT4Rec \cite{songNGAT4RecNeighborawareGraph2021}, which replaces the learnable linear transformation in vanilla GAT with cosine similarity to adapt to recommendation scenarios.
Most works adopt transformer architectures, adapting transformers to graph structures and recommendation scenarios through specialized positional encoding.
However, the computational complexity of global attention in transformers is mismatched with the data scale of recommendation tasks, often necessitating sampling \cite{chenSIGformerSignawareGraph2024}, masking \cite{chenMaskedGraphTransformer2024}, or approximation strategies \cite{chenRankformerGraphTransformer2025} to reduce time and space consumption.
More fundamentally, these GAT- and Transformer-based approaches inherit attention mechanisms designed for general machine learning tasks,  typically relying on monotonically increasing activation functions such as ReLU and Softmax, which fail to provide the sharp, ranking-aware distinctions near the top-K boundary where recommendation decisions are actually made.

\subsection{Graph Attention Networks}

Graph Attention Networks (GATs) \cite{velickovicGraphAttentionNetworks2017} revolutionized graph neural networks by introducing attention mechanisms to adaptively weight neighbor contributions during message passing. The original GAT computes attention coefficients through a learnable linear transformation, allowing each node to focus on the most relevant neighbors. This breakthrough inspired numerous extensions: GATv2 \cite{brodyHowAttentiveAre2021} identified and fixed the static attention problem in the original GAT by modifying the attention computation order; SuperGAT \cite{kimHowFindYour2020} introduces self-supervised edge attention to improve attention quality.

Recent work has applied attention mechanisms to graphs in various fields, including heterogeneous graph analysis \cite{wangHeterogeneousGraphAttention2019,huHeterogeneousGraphTransformer2020}, traffic flow prediction \cite{guoAttentionBasedSpatialtemporal2019, zhengGMANGraphMultiattention2020}, molecular prediction \cite{xiongPushingBoundariesMolecular2020, yingTransformersReallyPerform2021}, and so on.
These methods typically modify the attention mechanism by incorporating domain knowledge into the attention function, normalization schemes, and aggregation strategies.
In the recommendation domain, NGAT4Rec \cite{songNGAT4RecNeighborawareGraph2021} adapts its attention function for graph-based recommendations. And in the broader recommendation field, several works have explored graph attention network methods with external knowledge, such as KGAT \cite{wangKGATKnowledgeGraph2019} for knowledge graph-based recommendations, DGRec \cite{songSessionbasedSocialRecommendation2019} for session-based recommendations, and MGAT \cite{taoMGATMultimodalGraph2020} for multimodal recommendations. However, all these methods fundamentally inherit the attention design from general graph learning tasks, using monotonically increasing attention weights with respect to similarity and aggregation schemes optimized for traditional graph tasks, resulting in a critical mismatch between these methods and the top-K ranking objective of recommendation.

%% file: content/sec6-Conclusions.tex
\section{Conclusions}
We present \ours, a novel graph attention network architecture derived from optimizing the $\text{Precision}@K$ metric. Our key insight is that top-K recommendation fundamentally requires distinguishing items near the relevance boundary, which we achieve through a learnable threshold mechanism and a bandpass activation function. By establishing a direct mathematical connection between ranking metrics and neural architectures, \ours addresses the long-standing mismatch between graph-based recommendation models and their evaluation objectives. Extensive experiments demonstrate that our principled approach not only achieves state-of-the-art performance across multiple datasets but also provides interpretable insights into user preference patterns through the learned threshold dynamics. This work opens new avenues for metric-driven architecture design in recommendation systems, suggesting that future advances may benefit from similar first-principles approaches that align model design with task-specific objectives.

%% file: content/sec7-Acknowledgments.tex
This work is supported by the Zhejiang Province “JianBingLingYan+X” Research and Development Plan (2025C02020). We thank the reviewers for their valuable and insightful suggestions that improve the paper.

%% file: content/sec9-Appendices.tex
\section{Appendices}

\subsection{Derivation of Layer-wise Update} \label{app: derivation-layer}

First, we split $\mathcal L_{Pre@K} $ into two terms:
\begin{align}
& \mathcal{J}_{Pre@K} =\mathcal J_{TopK}+\lambda \mathcal J_{reg} \\
& \mathcal J_{TopK} = \sum_{u \in \mathcal{U}} \sum_{i \in \mathcal{N}_u} \frac{\sigma(s_{ui} - \beta_u^K)}{\sqrt{d_ud_i}}   = \sum_{(u,i) \in \mathcal{D}}  \frac{\sigma(s_{ui} - \beta_u^K)}{\sqrt{d_ud_i}} \\
& \mathcal J_{reg} =-\|\mathbf{Z}\|_2^2
\end{align}

The partial derivative of $\mathcal J_{TopK}$ with respect to $s_{ui}$ is: 
\begin{align}
\frac{\partial \mathcal J_{TopK}}{\partial s_{ui}}=\sum_{(u,i) \in \mathcal D} \frac{\sigma'(s_{ui} - \beta_u^K)}{\sqrt{d_ud_i}}
\end{align}
where $\sigma'(x)=\frac{1}{(1+e^{-x})(1+e^{x})}$ is the derivative of the Sigmoid function $\sigma(x)=\frac{1}{1+e^{-x}}$.

With$\frac{\partial s_{ui}}{\partial {\mathbf z}_u}={\mathbf z}_i$ and $\frac{\partial s_{ui}}{\partial {\mathbf z}_i}={\mathbf z}_u$, we obtain:
\begin{align}
\frac{\partial \mathcal J_{TopK}}{\partial {\mathbf z}_u}=\frac{\partial \mathcal J_{TopK}}{\partial s_{ui}}\frac{\partial s_{ui}}{\partial {\mathbf z}_u}=\sum_{i \in \mathcal N_u} \sigma'({\mathbf z}_u^T{\mathbf z}_i-\beta_u^K){\mathbf z}_i \label{eq: partial topk u} \\
\frac{\partial \mathcal J_{TopK}}{\partial {\mathbf z}_i}=\frac{\partial \mathcal J_{TopK}}{\partial s_{ui}}\frac{\partial s_{ui}}{\partial {\mathbf z}_i}=\sum_{u \in \mathcal N_i} \sigma'({\mathbf z}_u^T{\mathbf z}_i-\beta_u^K){\mathbf z}_u
\end{align}

The partial derivative of $\mathcal J_{reg} $ with respect to $\mathbf z_u$ and $\mathbf z_i$ are: 
\begin{align}
\frac{\partial\mathcal J_{reg}}{\partial \mathbf z_u}=-\mathbf z_u; \frac{\partial\mathcal J_{reg}}{\partial \mathbf z_i}=-\mathbf z_i
\label{eq: partial reg}
\end{align}

Integrating Eq.\eqref{eq: partial topk u}-\eqref{eq: partial reg}, we obtain:
\begin{align}
\frac{\partial \mathcal J_{Pre@K}}{\partial \mathbf z_u}=\frac{\partial \mathcal J_{TopK}}{\partial \mathbf z_u}+\lambda \frac{\partial \mathcal J_{reg}}{\partial \mathbf z_u}=\sum_{i \in \mathcal N_u} \sigma'({\mathbf z}_u^T{\mathbf z}_i-\beta_u^K){\mathbf z}_i-\lambda \mathbf z_u \label{eq: partial pre u}\\
\frac{\partial \mathcal J_{Pre@K}}{\partial \mathbf z_i}=\frac{\partial \mathcal J_{TopK}}{\partial \mathbf z_i}+\lambda \frac{\partial \mathcal J_{reg}}{\partial \mathbf z_i}=\sum_{u \in \mathcal N_i} \sigma'({\mathbf z}_u^T{\mathbf z}_i-\beta_u^K){\mathbf z}_u-\lambda \mathbf z_i \label{eq: partial pre i}
\end{align}

Substituting Eq.\eqref{eq: partial pre u}-\eqref{eq: partial pre i} into Eq.\eqref{eq: gradient descent}, we obtain:
\begin{align}
\mathbf z_u^{(l+1)}=(1-\tau\lambda)\mathbf z_u^{(l)}+\tau\sum_{i \in \mathcal N_u}\frac{{\sigma'\left(({\mathbf z}_u^{(l)})^T{\mathbf z}_i^{(l)}-\beta_u^{(l)}\right)}}{\sqrt{d_ud_i}}\mathbf z_i^{(l)} \\
\mathbf z_i^{(l+1)}=(1-\tau\lambda)\mathbf z_i^{(l)}+\tau\sum_{u \in \mathcal N_i}\frac{\sigma'\left(({\mathbf z}_u^{(l)})^T{\mathbf z}_i^{(l)}-\beta_u^{(l)}\right)}{\sqrt{d_ud_i}}\mathbf z_u^{(l)}
\end{align}
The value range of $\sigma'(\cdot)$ is $[0,\frac{1}{4}]$. To maintain numerical scale and facilitate the expression of subsequent formulas, we denote $\omega(\cdot)=4\sigma'(\cdot)$, obtaining Eq.\eqref{eq: after omega u}-\eqref{eq: after omega i}.

\subsection{Experimental Parameter Settings}
For our \ours, we employ the BPR loss function and the Adam optimizer, and perform grid search to tune the hyperparameters. Specifically, we set the hidden embedding dimension $d$ to $64$, following previous work \cite{heLightGCNSimplifyingPowering2020, leeRevisitingLightGCNUnexpected2024, chenRankformerGraphTransformer2025}. The learning rate is searched in the range of $\{0.1, 0.01, 0.001\}$, and the weight decay is searched in the range of $\{0, 1e^{-8}, 1e^{-4}\}$. The number of layers $L$ is chosen in the range of $\{1, 2, 3, 4, 5\}$.